\documentclass[prb,twocolumn,showpacs,preprintnumbers]{revtex4}
\usepackage{graphicx}
\usepackage{dcolumn}

\begin{document}

\title{How Magnetic Field Enters Heat Current:
\\ Application to Fluctuation Nernst Effect}

\author{ A. Sergeev}
\email{asergeev@eng.buffalo.edu} \affiliation{ Research
Foundation, University at Buffalo, Buffalo, New York 14260}
\author{M. Yu. Reizer} \affiliation{5614 Naiche Rd.
Columbus, Ohio 43213}
\author{V. Mitin} \affiliation{Electrical Engineering
Department, University at Buffalo, Buffalo, New York 14260}


\begin{abstract}
A problem of the definition of the heat transported in
thermomagnetic phenomena has been well realized in the late
sixties, but not solved up to date.  Ignoring this problem,
numerous recent theories grossly overestimate  the thermomagnetic
coefficients in strongly interacting systems. Here we develop a
gauge-invariant microscopic approach, which shows that the heat
transfer should include the energy of the interaction between
electrons and a magnetic field. We also demonstrate that the
surface currents induced by the magnetic field transfer the charge
in the Nernst effect, but do not transfer the heat in the
Ettingshausen effect. Only with these two modifications of the
theory, the physically measurable thermomagnetic coefficients
satisfy the Onsager relation. We critically revised the Gaussian
fluctuation model above the superconducting transition and show
that the gauge invariance uniquely relates thermomagnetic
phenomena in the Fermi liquid with the particle-hole asymmetry.
\end{abstract}

\pacs{PACS   numbers: 71.10.-w} \maketitle

\section{Introduction}
It is well known that in the electric field, ${\bf E} =-\nabla
\phi $, and magnetic field ${\bf H}$, the energy of a system of
charged particles with coordinates ${\bf r}_\alpha$ and momentums
${\bf p}_\alpha$  is \cite{1}
\begin{eqnarray}\label{energy}
E = \sum_\alpha {p_\alpha^2 \over 2m} +  \sum_\alpha e \phi({\bf
r}_ \alpha) + \sum_\alpha {e\over 2mc} ({\bf r}_\alpha \times {\bf
p}_\alpha) \cdot {\bf H}.
\end{eqnarray}
Considering the heat transfer, it is important to realize which
terms in the above equation should be associated with the thermal
energy. The uniform electric field accelerates all particles in
the same way, while the effect of the magnetic field depends on
the particle state. Therefore, it is reasonable to assume that the
electric term (the second term) contributes to the potential
energy, while the magnetic term (the third term) provides
contribution to the thermal energy. Of course, the above
consideration cannot be considered as a proof. Moreover, there is
no regular methods for identification of thermal energy.
Correctness of our assumption will be verified through the Onsager
relation between the Ettingshausen and Nernst coefficients, which
describe the heat and electric currents initiated by the electric
field and temperature gradient, correspondingly. We will show that
the magnetic term overlooked in previous works plays a crucial
role and restores the gauge invariance. While in accordance with
van Leeuwen's theorem \cite{1} the magnetic contribution to the
heat transfer does not allow a classical interpretation, it is
critically important for consistent quantum description of
thermomagnetic transport.

The definition of the heat current and the Onsager relation in the
magnetic field are long standing problems, which were attracted
significant attention in the late sixties in the context of
thermomagnetic vortex transport in the type-II superconductors
\cite{2,3}. Maki \cite{2} has noted that the phenomenological
\cite{4} and microscopic \cite{5} descriptions of thermomagnetic
phenomena violate the third law of thermodynamic and the Onsager
relation. All attempts to resolve this enigma were based on
thermodynamic treatment. Various corrections suggested to the heat
current are expressed in terms of the equilibrium magnetization
currents \cite{2,3}. As we will show, the magnetization currents
do not transfer the heat and the thermodynamic approach cannot
solve this problem.

Recently, much efforts have been dedicated to understanding of
thermomagnetic effects in high-$T_c$ superconductors \cite{6,7,8}.
Variety of phenomenological concepts were suggested to explain
anomalously large Nernst signal above $T_c$ \cite{9}. In many
papers \cite{10,11,12,13} the Nernst effect is attributed to the
motion of vortices created by phase fluctuations. Microscopic
models \cite{14,15} are based on the Gaussian-Fluctuation Theory
(GFT), which includes both module and phase fluctuations, but is
applicable only near $T_C$ \cite{16,17,18,19} (for a review see
\cite{20}). However, the theories related to vortices are based on
the results of Refs. \cite{4} and \cite{5}, which inconsistency
with basic principles was noted in \cite{2,3}. GFT and other
microscopic models also ignore corrections to the heat current.

While the large Nernst coefficient was obtained in all above
models, there is still not much understanding of how
superconducting fluctuations could lead to this effect. In the
model of noninteracting electrons, which describes well ordinary
metals, the Nernst coefficient is small, because it is
proportional to the square of the particle-hole asymmetry (PHA)
\cite{21}. In this case, the Nernst coefficient combines PHAs of
the thermoelectric, $\eta$, and Hall, $\sigma_{xy}$, coefficients:
$N_n \simeq \eta \cdot \sigma_{xy}$. The interelectron interaction
cannot change PHA of $\eta$ and $\sigma_{xy}$ \cite{22}. According
to current points of view \cite{9,14,15,16,17,18,19}, the
interaction in the Cooper channel leads to the Nernst coefficient
in the zeroth order in PHA. This effect is $(\epsilon_F/T)^2$
times larger than the corresponding correction to $\eta \cdot
\sigma_{xy}$ ($\epsilon_F$ - is the Fermi energy). However,
experiments with ordinary superconductors, such as Nb, Al, Sn, do
not support this prediction.  This contradiction is a consequence
of a number of misconceptions at microscopic level.

The microscopic formalism based on the Kubo method allows one to
calculate the Ettingshausen coefficient $\Upsilon$, which
describes the heat flux induced by crossed electric and magnetic
fields, ${\bf j}^h =-\Upsilon [{\bf E} \times {\bf H}]$. The
temperature gradient cannot be directly introduced in the Lagrange
formalism of the Kubo approach. Therefore, the Nernst coefficient,
which relates the electric current with the temperature gradient,
${\bf j}^e = N[\nabla T \times {\bf H}]$, is found from the
Onsager relation,
\begin{eqnarray}\label{Ons}
N=\Upsilon/T.
\end{eqnarray}
The coefficient $\Upsilon$ is calculated for the infinite sample,
while the Onsager relation is only valid for a finite sample
\cite{23}, where surface currents should be taken into account.

If magnetization depends on temperature, the temperature gradient
generates surface magnetization currents, which provide important
contribution to the charge transfer in the Nernst effect
\cite{23}. Several recent works \cite{18,19,24} state that surface
magnetization currents also play an important role in the heat
transfer and the Ettingshausen coefficient calculated for the
infinite sample, $\delta \Upsilon_{inf}$, should be corrected due
to the surface heat currents: $\delta \Upsilon = \delta
\Upsilon_{inf} - c\mu$. According to \cite{16,18,19}, GFT leads to
the bulk coefficient, which is of the same order as the correction
due to surface currents: $\delta \Upsilon_{inf}= 3/2 \ c\mu$,
where $\mu$ is the fluctuation magnetic susceptibility \cite{20},
\begin{eqnarray}\label{mu}
 \mu &=& -
{e^2 T \over 6 \pi c^2} \times \cases{
   2 \alpha/\eta
    & for 2D, \cr
    \sqrt{\alpha/\eta}
    & for 3D, \cr}
\end{eqnarray}
$\eta=(T-T_c)/T_c$ and $\alpha = (T-T_c)[\xi (T)]^2 $, $\xi (T)$
is the coherence length.

In the current paper we reconsider basics of the microscopic
description of the thermomagnetic effects. This work will address
the following questions: (1) Is the heat current operator modified
in the presence of a magnetic field? (2) Do surface magnetization
currents transfer the heat? (3) How do thermomagnetic coefficients
satisfy the Onsager relation? (4) Finally, can the interelectron
interaction change the PHA of the Nernst and Ettingshausen
coefficients? While our results are quite general, we specify them
for GFT, which will be critically revised. In particular, we show
that $\delta \Upsilon \sim T \delta N \sim (T_c/\epsilon_F)^2
\mu$.

\section{Ettingshausen coefficient: Kubo method}
In the gauge ${\bf H}= i[{\bf k}_H \times {\bf A}_H ]$ and ${\bf
E}= -i {\bf k}_E \phi $, contributions of static electric and
magnetic fields to the energy of a charged particle are easy
separated,
\begin{eqnarray}\label{Energy}
 \tilde{E}={({\bf p}+e{\bf A}_H/c)^2 \over 2m} +e\phi
= \xi_p + {e\over c} ({\bf v}\cdot {\bf A}_H) + e\phi +\mu_0,
\end{eqnarray}
where $\xi_p = p^2/2m - \mu_0$, $\mu_0$ is the chemical potential.
Let us first show the importance of the second (magnetic) term for
noninteracting electrons \cite{25}. Thermal energy is counted from
the electro-chemical potential, so only the first and the second
terms contribute to the heat current.  The corresponding diagrams
for the Ettingshausen coefficient are shown in Fig. 1. The diagram
(a) has a form of the Hall diagram \cite{26}, where the electric
current operator, $e {\bf v}$, is replaced by the heat current, $
\xi_p {\bf v}$ . The Hall effect is proportional to PHA, i.e. the
corresponding integrant is an odd function of $\xi_p$ and a
nonzero Hall coefficient is obtained after expansion of all
parameters in $\xi_p/\epsilon_F$ near the Fermi surface. With the
heat current operator $ \xi_p {\bf v}$, the integrant becomes an
even function and gives a nonzero Ettingshausen coefficient
without expansion. However, this large contribution is canceled by
the diagram (b). The well-known expression for $\Upsilon$ in a
system of noninterating electrons is obtained in the second order
in PHA, when both integrants (a) and (b) are expanded in
$(\xi_p/\epsilon_F)^2$ (for details see an Appendix in \cite{25}).

Now we consider electrons interacting in the Cooper channel. The
energy current operator in this system \cite{27} is easily
expanded to include external fields,
\begin{eqnarray}\label{Jh}
{\hat {\bf J}}^\epsilon = \sum_{\bf p} {\bf v} \xi_p  a_{\bf p}^+
a_{\bf p} + \sum_{\bf p} {e{\bf v}\over c} ({\bf v}{\bf A}_H)
a_{\bf p}^+a_{\bf p} + \sum_{\bf p} {\bf v} e\phi  a_{\bf p}^+
a_{\bf p} \nonumber \\ + \sum_{\bf p} {\bf v} \mu \ a_{\bf p}^+
a_{\bf p} - {\lambda /2} \sum_{{\bf p},{\bf p}',{\bf p}''} ({\bf
v}+{\bf v}') \ a_{{\bf p}+{\bf p}'-{\bf p}''}^+ a_{{\bf p}''}^+
a_{{\bf p}'} a_{\bf p} \nonumber \\ + \sum_{{\bf p},{\bf p}',{\bf
R}_i} ({\bf v}+{\bf v}') U_{imp}({\bf R}_i) \ \exp [i({\bf p}+{\bf
p}') R_i] \ a_{\bf p}^+ a_{{\bf p}'} \ , \ \ \
\end{eqnarray}
here $a_{\bf p}^+$ and $a_{\bf p}$ are the electron creation and
annihilation operators, $\lambda$ is the interaction constant in
the Cooper channel, and $U_{imp}$ is the impurity potential. The
first and the forth terms in the Eq. \ref{Jh} describes the energy
flux of noninteracting electrons without fields. The second and
third terms are due to electron interaction with magnetic and
electric fields. Two last terms are due to the electron-electron
and electron-impurity interactions. These two terms generate
diagram blocks (Aslamazov-Larkin blocks) proportional to PHA, but
contributions of all blocks with PHA cancel each other \cite{27}.

The thermal energy is defined as the electron energy counted from
the electro-chemical potential $e\phi+\mu$ and, therefore, the
third and the forth terms in Eq. \ref{Jh} do not contribute to the
heat current. Thus, calculating the heat current we should take
into account only two heat current vertices $\gamma^h_1$ and
$\gamma^h_2$ corresponding to the first (kinetic) and second
(magnetic) terms in Eq. \ref{Jh}.

Two leading terms in the heat current operator generate two
diagrams, which describe $\Upsilon$ in the Aslamazov-Larkin (AL)
approximation (see Fig. 2). The wavy lines correspond to the
fluctuation propagator \cite{20,27},
\begin{eqnarray}
L^{R,A}(q,\omega) &=& \bigl(\lambda^{-1}-P^{R,A}(q,\omega)
\bigr)^{-1}, \label{L1} \\
 P^{R,A}(q,\omega)&=& -{\nu \over 2} \biggl
(\ln{{2\ C_\gamma \omega_D \over \pi T}} -\alpha q^2 \pm {i\pi
\omega \over 8T} +\gamma \omega \biggr),\label{P2} \ \
\end{eqnarray}
where $P(q,\omega)$ is the polarization operator, $\nu$ is the
electron density of states, $\omega_D$ is the Debye frequency, and
$C_\gamma$ is the Euler constant. The last term in Eq. \ref{P2} is
proportional to PHA \cite{27,28}.

The AL blocks ${\bf B}^{e,h,H}$ presented in Fig. 2 are built from
electron Green functions and vertices $ \gamma^{e,h,H} $ (see Tab.
1). The left blocks ${\bf B}^h_1$ and ${\bf B}^h_2$ in Figs. 2.a
and 2.b are blocks with heat current vertices $\gamma^h_1$
(kinetic) and $\gamma^h_2$ (magnetic). The right block ${\bf B}^e$
in both diagrams includes the electric current vertex $\gamma^e=
e{\bf v}\cdot{\bf e}_E $, ${\bf e}_E = {\bf E}/E $. Block ${\bf
B}^H$ includes the magnetic vertex $\gamma^H = (e/c){\bf
v}\cdot{\bf A}_H$. Results of calculation are summarized in Table
I.  AL blocks are obtained by insertion vertices $ \gamma$ into
the polarization operator and can be expressed through
$\nabla_{\bf q} P^R({\bf q},0)$. Blocks ${\bf B}^e$, ${\bf B}^H$,
and ${\bf B}^h_1$ are well known \cite{20}. The block ${\bf
B}^h_1$ describing the heat current in the absence of magnetic
field has been calculated in \cite{27} (see also \cite{18,19}).
Here we introduce ${\bf B}^h_2$, which is based on the electron
vertex $ \gamma^h_2 $ and describes the magnetic correction to the
heat current.

The first AL diagram (Fig. 2.a) was investigated in Ref. \cite
{19} and its contribution is $\delta \Upsilon_{inf}^{(1)} = 3/2 \
c\mu$. The same result has been obtained in the time dependent
Ginzburg-Landau formalism (TDGL) \cite{16,18,20}.

Contribution of the second AL diagram (Fig. 2.b) is
\begin{eqnarray}\label{D2}
\Upsilon_{inr}^{(2)}H = \Im  \int {d{\bf q} \over (2\pi ) ^n}
{d\omega \over 2\pi} \ {{\bf B}^h_2  {\bf B}^e\over 2 \Omega}
(L^C_+L^A_- +L^R_+ L^C_-), \
\end{eqnarray}
where $L^C=\coth(\omega/2T)(L^R-L^A)$, $L_\pm$ is used for $L({\bf
q} \pm {\bf k}/2, \omega \pm \Omega/2)$, and $n$ is the system
dimensionality with respect to the coherence length $\xi (T)$.
Expanding the integrant to the linear order in $\Omega$ and {\bf
k} and calculating the integrals over $\omega$ and ${\bf q}$, we
find that the contribution of the second diagram,
$\Upsilon_{inr}^{(2)}$ cancels completely the contribution of the
first one.

Thus, without PHA the Ettingshausen effect is absent, $\delta
\Upsilon=0$. To get nonzero result, we should expand the
fluctuation propagator (Eq. \ref{L1}) up to the second order in
PHA. Expanding the polarization operator (Eq. \ref{P2}) to the
second order in $\gamma \omega$ we get
\begin{eqnarray}\label{Ett}
{\delta \Upsilon \over T} =  \delta N = - {5 e^2  \over 4\pi
c}\biggl({8T \gamma\over \pi }\biggr)^2  \cases{
    2 \alpha/\eta
    & for 2D, \cr
    \sqrt{\alpha/\eta}
    & for 3D; \cr}
 \end{eqnarray}
where ${\displaystyle \gamma = {1 \over 2 \epsilon_F} {\partial
\ln \nu \over
\partial \ln \epsilon_F}  \ \ln {2 C_\gamma \omega_D \over \pi
T_c}} \ $ \cite{28}. Thus, thermomagnetic coefficients in the
fluctuation region are proportional to $(T/\epsilon_F)^2$.

Summarizing results of this section, we would like to note that
the total heat current operator of the fluctuating pairs in the
magnetic field,
\begin{eqnarray}\label{Bh}
{\bf B}^h={\bf B}^h_1+{\bf B}^h_2= \omega \nu\alpha \ [{\bf q}
+(2e/c){\bf A}_H],
 \end{eqnarray}
may be considered as the gauge-invariant extension of the operator
${\bf B}^h_1$ without ${\bf H}$. This is a key point, because the
further calculations of diagrams in Fig. 2 are similar to that for
noninteracting electrons (Fig. 1): the kinetic and magnetic terms
in ${\bf B}^h$ generate two diagrams, which cancel each other in
the zeroth order in PHA.

In the above calculations the interaction with the magnetic field
has been included in the heat current. Note that Eq. \ref{Bh} can
also be derived in another, more formal approach, where the
magnetic field is initially included in electron states. Without
magnetic field, the thermoelectric coefficient is described by the
AL diagram with the heat and electric current operators, ${\bf
B}^h_1$ and ${\bf B}^e$ \cite{20,27}. In the magnetic field, the
momentum of the Cooper pair is given by ${\bf q} + 2e{\bf A}_H/c$,
and the polarization operator has a form: $P({\bf q} + 2e{\bf
A}_H/c, \ \omega)$ \cite{20}. Calculating the thermomagnetic
response, all blocks of the diagram should be expanded in ${\bf
A}_H$. Expanding polarization operators in the fluctuation
propagators, we obtain the first diagram for the thermomagnetic
coefficient \cite{19,20}. Expanding the heat current block ${\bf
B}^h_1 = \omega \ \nabla_{\bf q} P^R({\bf q}+ 2e{\bf A}_H/c)$, we
immediately obtain the block ${\bf B}^h_2$, which forms the second
diagram. This magnetic term has been lost in all previous works
\cite{16,17,18,19}. Using Eq. 10 as the heat current operator for
fluctuating pairs, our results can be also obtained in TDGL
formalism.

\section{Nernst coefficient: Quantum transport equation}
To investigate a response of the electron system to $\nabla T$,
one can use the quantum transport equation. Previously we adapted
this method to calculations of the themoelectric and Hall
coefficients in GFT \cite{25,29}. In this approach the electric
current is given by
\begin{equation}\label{Je}
j^e  = e\nu \alpha \int { d{\bf q}\over (2 \pi)^n}  \ {d\omega
\over (2 \pi)} \ {\bf q} \ {\rm Im}  \ \delta L^C ({\bf q},
\omega),
\end{equation}
where $\delta L^C ({\bf q}, \omega)$ is the nonequilibrium
correction to the fluctuation propagator. In the equilibrium, $L^C
= L^R P^C L^A$, where the Keldysh component of the polarization
operator $P^C = i \pi \nu/4$ at $T-T_c \ll T_c$.

Calculation of $\delta L^C ({\bf q}, \omega)$ for the Nernst
coefficient is analogous to its calculation for the Hall effect
\cite{29}. The nonequilibrium effects are taken into account by
the $\nabla T$ and ${\bf H}$-Poisson brackets between polarization
operators: $\lbrace P_1,P_2\rbrace_T = \nabla T (\partial_T P_1
\cdot
\partial_q P_2 - \partial_T P_2 \cdot \partial_q P_1) $
and $ \lbrace P_1,P_2 \rbrace_H = (e / c){\bf H}\cdot [\partial_q
P_1  \times \partial_q P_2] $, where $\partial_T P = {\partial P /
\partial T}$ and $\partial_q P = {\partial P /
\partial{\bf q}}$
 \cite{25,27,29}. Therefore, the only difference in
calculations of the Hall and Nernst coefficients is that the
derivatives $
\partial P / \partial \omega$ in Hall coefficient (${\bf
E}$-Poisson bracket \cite{29}) should be replaced by the
derivatives $\partial P / \partial T $ in the Nernst coefficient.

To get the vector product ${\bf H}\times \nabla T$, the ${\bf
H}$-bracket should include the same polarization operator as the
$\nabla T$-bracket. Thus, in the first order in ${\bf H}\times
\nabla T$ the nonequilibrium fluctuation propagator $\delta L^C
({\bf q}, \omega)$ is described by the diagrams shown in Fig. 3.
Finally, calculating $\delta L^C ({\bf q}, \omega)$ and the
corresponding Nernst current in the interior of the sample (Eq.
\ref{Je}), we find
\begin{eqnarray}\label{Ninf}
\delta N_{inf} = \delta N+{e^2 \over 3 \pi c} \biggl( {\alpha
\over \eta^2} -{\alpha \over \eta}- {\alpha \over \eta} {\partial
\alpha \over \partial T}\biggl),
\end{eqnarray}
where $\delta N$ is equal to the term, which was calculated in the
previous section from the Onsager relation (see Eq. \ref{Ett}).
Thus, in the infinite sample or in the interior of the finite
sample, the Nernst coefficient consists of two terms. The second
term in Eq. \ref{Ninf} has the zeroth order in PHA and violates
the Onsager relation. As it will be shown in the next section, in
the finite sample this term is canceled by the contribution of the
magnetization currents.

\section{Onsager relation in magnetic field}
The above results, Eqs. \ref{Ett} and \ref{Ninf}, have been
calculated for the infinite sample. Referring to \cite{23}, recent
works \cite{18,19,24} state that for a finite sample both
coefficients should be corrected due to charge and heat transfer
by surface magnetization currents. Here we show that the
magnetization currents contribute only to the charge transfer, and
the results of \cite{23} have been misinterpreted.

The electric magnetization current ${\bf j}^e_{mag}$ in the
potential relief $\phi ({\bf r})$ transfers the energy flux ${\bf
j}^\epsilon_{mag}=\phi{\bf j}^e_{mag}$ (Eq. 37 in \cite{23}).
Using ${\bf j}^e_{mag} = c \mu k^2 {\bf A}_H $, we get
\begin{eqnarray}\label{magcur}
{\bf j}^\epsilon_{mag}= c\mu[{\bf H}\times{\bf E}].
 \end{eqnarray}
This term is erroneously attributed to the heat flux
\cite{18,19,20,24}. As we discussed, the electric potential $\phi$
and the corresponding vertex $\gamma^\phi= e{\bf v} \phi$ do not
contribute to the heat current, because the thermal energy should
be counted from the electro-chemical potential.

In the interior of the sample the electric current consists of the
transport and magnetization components, ${\bf j}^e_{inr} = {\bf
j}^e_{tr}+{\bf j}^e_{mag}$. The magnetization component is
\cite{23}
\begin{eqnarray}\label{jem}
{\bf j}^e_{mag} = c \ {\partial \mu \over \partial T} \ (\nabla T
\times {\bf H} ).
\end{eqnarray}
The magnetization currents are divergence-free. The total
magnetization current through the sample cross-section must be
zero, i.e. the bulk magnetization currents are canceled by the
surface currents. Therefore, the Nernst coefficient measured in
the finite sample is determined by the transport currents: $N={\bf
j}^e_{tr}/[\nabla T \times {\bf H}]$ \cite{23}. The Nernst
coefficient in the infinite sample (Eq. \ref{Ninf}) is associated
with the bulk current in the finite sample, $N_{inf}={\bf
j}^e_{inr}/[\nabla T \times {\bf H}]$. Using Eq. \ref{jem}, we get
\begin{eqnarray}\label{Ons1}
\delta N = {{\bf j}^e_{inr} -{\bf j}^e_{mag} \over [\nabla T
\times {\bf H}]} = \delta N_{inf} - c {\partial \mu \over
\partial T}.
\end{eqnarray}
Taking into account Eq. \ref{mu}, we see that the second term in
the last equation, $c\partial \mu/\partial T $, cancels completely
the second term in $\delta N_{inf}$ (Eq. \ref{Ninf}). The rest is
equal to $\delta N$, which satisfies the Onsager relation $\delta
N= \delta \Upsilon/ T$.

Note that, if in contradiction to our results, the surface
magnetization currents provide the heat transfer, this effect
could be found in transport measurements. In the Gorbino disc
geometry with the magnetic field perpendicular to the disc and the
circular inductive electric field in the plane (Fig. 4), the heat
current in the radial direction does not contain the surface
components, which were predicted for the standard parallelepiped
geometry in \cite{18,19,24}. According to our results, both
experiments will give the same results. The surface electric
currents generated by $\nabla T$ are very significant (see Eq.
\ref{jem}). However, they cannot be experimentally separated from
the interior currents, because contrary to the circular electric
field, the circular temperature gradient does not exist. This
difference between ${\bf E}$ and $\nabla T$ is reflected in the
asymmetry of the Nernst and Ettingshausen coefficients calculated
for the infinite sample.

\section{Particle-hole asymmetry in thermomagnetic effects}
Now we show that in the general case, the interelectron
interaction cannot change PHA requirements for $N$ and $\Upsilon$,
i.e. the thermomagnetic coefficients are always proportional to
the square of PHA.

Assuming that electron scattering from impurities is the main
mechanism of the momentum relaxation, it is easy to see
\cite{25,27,29} that the magnetic field and temperature gradient
enter into the transport equation formalism through the
distribution functions of noninteracting electrons and the Poison
brackets. In fact, the terms proportional to $\nabla T \times {\bf
H}$ can appear in three different ways \cite{25}: (${\it a}$)
through the Nernst nonequilibrium distribution function of
noninteracting electrons, $({e\tau^2 / cm}) \ {\bf v}\cdot[\nabla
T \times{\bf H}] \ ({\partial S /
\partial T})$, (${\it b}$) through the ${\bf H}$-Poisson bracket
that involves the nonequilibrium distribution function under the
temperature gradient, $-e \tau ({\bf v} \cdot {\bf E}) (\partial
S/\partial T)$, and, finally, (${\it c}$) due to double, $\nabla
T$ and  ${\bf H}$, Poisson brackets. It is evident that ${\it a}$
and ${\it b}$-type terms have already includes the Hall PHA, which
is proportional to $(\partial v/
\partial p) = 1/m$. The ${\it c}$-type terms in the form of the
double Poisson brackets describe the AL process, which has been
investigated above. As we have seen, the AL diagram gives the
contribution in the zeroth order in PHA, however, this
contribution is canceled by the contribution of the surface
magnetization currents. Thus, we conclude that the interelectron
interaction can provide many-body thermomagnetic effects only in
the second order in PHA.

\section{Conclusions}
We have shown that the magnetic term in the Hamiltonian of charged
particles (Eq. \ref{energy}) should be associated with the thermal
energy. The corresponding term in the heat current operator (Eq.
\ref{Energy}) restores the gauge invariance and gives important
contribution to the Ettingshausen coefficient. We also found that
the surface magnetization currents do not contribute to the heat
current, but provide substantial contribution to the charge
transfer in the Nernst effect (Eq. \ref{Ons1}). Our
gauge-invariant scheme gives the thermomagnetic coefficients that
satisfy to the Onsager relation (see Eqs. \ref{Ons1} and
\ref{Ninf}). In the general case of the Fermi liquid with
particle-hole excitations, we conclude that the measured
thermomagnetic coefficients are always proportional to the square
of PHA. Any interaction by itself, i.e. without changing the
electron band structure or character of elementary excitations,
cannot provide large thermomagnetic effects.

The developed approach has been applied to effects of
superconducting fluctuations. We show that the gauge invariant
form of the heat current operator of fluctuating pairs is $\nu
\alpha \omega ({\bf q} + 2e{\bf A}_H/c)$. The second (magnetic)
term missed in previous publications plays an important role: as
in the case of noninteracting electrons, the corresponding diagram
 cancels completely the large, zeroth-order in PHA term in the
Ettingshausen coefficient.  We also show that the Nernst
coefficient in the infinite superconducting sample consists of an
anomalously large, the zeroth order in PHA term (Eg. \ref{Ninf}).
However, in the finite sample, this term is cancelled by the the
surface magnetization currents (Eq. \ref{Ons1}) and thermoelectric
coefficients satisfy the Onsager relation. Our results for $\delta
N $ are different by a huge factor of $(\epsilon_F / T)^2$ from
the previous works \cite{16,17,18,19}, which claim that the
attractive interaction in the Cooper channel provide
thermomagnetic transport without PHA at all. We can also easily
rebuff this claim, if we just change a sign of the interaction
constant and consider the repulsive interaction in the Cooper
channel. As known, in ordinary metallic films this interaction
results in corrections to conductivity, which are a factor of
$(\epsilon_F \tau)^{-1}$ smaller than the conductivity of
noninterating electrons \cite{30}. If the statement of Refs.
\cite{16,17,18,19} is correct, the thermomagnetic effects in
ordinary metals would be $(\epsilon_F \tau)^{-1} (\epsilon_F/T)^2
\sim (\epsilon_F/T)/(T \tau)$ larger than predicted for
noninteracting electrons. Certainly this huge effect is not known.

Returning to the problem of high-$T_c$ superconductors, we should
note that most of the models, including fluctuation exchange
\cite{15}, 'cheap' vortices \cite{9}, and preformed pairs
\cite{14}, have treated the state with large Nernst effect as
natural extensions of GFT. In the light of our results, the
concept of this specific state which continuously evolved from GFT
due to fluctuations and additional interactions is far from
obvious. In this sense, the current work rather supports an idea
of the Nernst state as "a thermodynamic phase in itself with
characteristic properties specific to it as a fixed point,
distinct from those of a normal metal" \cite{12}.

Using GFT as an example, we have shown that the gauge-invariant
form of the heat current is critical for description of the
Ettingshausen effect and that the surface currents are important
for the Nernst effect. All other models of thermomagnetic
transport including vortex models should be reconsidered in
accordance with the formalism developed above.

We would like to acknowledge useful discussions with I. Aleiner,
A. Larkin, D. Livanov, A. Varlamov, and I. Ussishkin.

\appendix
\section{Method Kubo for Noninteracting electrons}

\centerline{\bf Appendix A}

In this appendix we present detailed calculations of the
Ettingshausen coefficient for noninteracting electrons using the
Kubo method.

According to Eq. 4, the heat current vertex for noninteracting
electrons in the magnetic field is given by
\begin{eqnarray}\label{A1}
\gamma^h= \gamma^h_1+\gamma^h_2 = \xi_p {\bf v} + (e/c) ({\bf v}
\cdot {\bf A})_H {\bf v}.
\end{eqnarray}
Two vortices
$\gamma^h_1$ and $\gamma^h_2$ create two diagrams shown in Fig. 1.
Solid lines in diagrams represent the electron Green functions,
\begin{eqnarray}\label{A2}
G^R_p = [G^A_p]^* = (\epsilon-\xi_p+i/2\tau)^{-1}.
\end{eqnarray}
To get the Ettingshausen coefficient proportional to ${\bf A}({\bf
k} \cdot {\bf E})$, one should expand the Green function $G({\bf
p}+ {\bf k}) $ in powers of $({\bf k}\cdot {\bf v})$. Then the
contribution of the first diagram is given by
\begin{eqnarray}\label{A3}
\Upsilon_1 = {e^2 \over c H}\int {d {\bf p} \over (2\pi)^n} {d
\epsilon \over 2\pi} { \partial S_0  \over
\partial \epsilon } \ \xi_p {\bf v} ({\bf v} \cdot
{\bf A}) ({\bf v} \cdot {\bf E}) ({\bf v} \cdot {\bf k}) I_1, \ \
\
\end{eqnarray}
where $S_0=-\tanh(\epsilon/2T)$, $n$ is the dimensionality of the
system, and the combination of the Green functions is
\begin{eqnarray}\label{A4}
I_1 = 2i G^A_p G^R_p {\rm Im} (G^A_p)^2= 2i \tau^2 {\rm Im}
(G^A_p)^2.
\end{eqnarray}
The contribution of the second diagram is
\begin{eqnarray}\label{A5}
\Upsilon_2= {e^2\over cH} \int {d {\bf p} \over (2\pi)^n} {d
\epsilon \over 2\pi} {\partial S_0 \over \partial \epsilon } \
{\bf v}({\bf v} \cdot {\bf A}) ({\bf v} \cdot {\bf E}) ({\bf v}
\cdot {\bf k})I_2, \ \ \
\end{eqnarray}
where the combination of the Green functions $I_3$ is
\begin{eqnarray}\label{A6}
I_2 = 2i G^A_p {\rm Im} (G^A_p)^2=2i\tau^2 {\rm Im}G^A_p.
\end{eqnarray}
Integration over angles of the electron momentum in Eqs. (A.3) and
(A.5) gives
\begin{eqnarray}\label{A7}
\int d \Omega_{\bf p}  \ {\bf v} ({\bf v} \cdot {\bf A}) ({\bf v}
\cdot {\bf k}) ({\bf v} \cdot {\bf E}) = {v^4 \over n(n+2)} {\bf
A} ({\bf k} \cdot {\bf E} ). \ \ \
\end{eqnarray}
Then the total contribution may be presented as
\begin{eqnarray}\label{A8}
\Upsilon &=& {i e^2 \over n(n+2) cH} \int {d \epsilon \over 2\pi}
d\xi_p {
\partial S_0 (\epsilon) \over
\partial \epsilon } \ {\bf A} ({\bf k} \cdot {\bf E}) \nonumber
\\
&\times& v^4 \tau^2 \nu \biggl(\xi_p {\rm Im} (G^A_p)^2 +{\rm Im}
G^A_p \biggr).
\end{eqnarray}
Without taking into account PHA the total contribution of two
diagrams goes to zero after integration over $\xi_p$, because
\begin{eqnarray}\label{A9}
\int d \xi_p \biggl( \xi_p (G^A_p)^2 +{\rm Im} G^A_p \biggr)
 =0.
\end{eqnarray}
Nonzero contribution arises from terms proportional to
$\epsilon^2$, thus we should expand all electron parameters near
the Fermi surface. For example for 3D conductor,
\begin{eqnarray}\label{A10}
v^4 \nu &=& v_0^4 \nu_0 \bigg[ 1+ {5\over 2}{\xi_p \over
\epsilon_F} +{15 \over 8}\bigg({\xi_p\over \epsilon_F} \bigg)^2+
...\bigg], \\ \tau^2 &=& \tau_0^2 \bigg[1 -{\epsilon \over
\epsilon_F} -\bigg({\epsilon \over \epsilon_F}\bigg)^2 +
...\bigg].
\end{eqnarray}
Taking into account terms proportional to the square of PHA, e.g.
$\xi^2/ \epsilon_F^2$ or $ \xi \epsilon /\epsilon_F^2$, we get
\begin{eqnarray}\label{A12}
\int d \xi_p \ v^4 \nu \tau^2 \biggl(\xi_p {\rm Im}[G^A(P)^2 +{\rm
Im}G^A(P)\biggr)\ \ \ \ \ \ \ \ \ \nonumber \\ = - \pi v_0^4 \nu_0
\tau_0^2 {5 \over 4} {\epsilon^2 \over \epsilon_F^2} = -\pi {5
\over 2} {v_0^2 \tau_0^2 \nu_0 \over m} {\epsilon^2 \over
\epsilon_F}. \ \ \
\end{eqnarray}
Substituting this result into Eq. (A.8), and performing
integration over $\epsilon$, we get the well-known result for the
Ettingshausen coefficient of noninteracting electrons,
\begin{eqnarray}\label{A13}
\Upsilon_{3D} =-{\pi^2\over 6}{T^2 \over
\epsilon_F}(\Omega\tau_0){\sigma_{xx}\over H}, \ \ \
\end{eqnarray}
where $\Omega=eH/mc$ is the cyclotron frequency and $\sigma_{xx}$
is the Drude conductivity. For 2D-conductors, the corresponding
relation between $\Upsilon_{2D}$ and two-dimensional conductivity
has an additional numeric factor of 2.

Thus, calculating the Ettingshausen coefficient of noninteracting
electrons, we demonstrated that the magnetic field should be taken
into account in the heat current vertex. The diagram with this
magnetic vertex in the heat current cancels the basic diagrams in
the zeroth order in PHA. The nonzero Ettingshausen coefficient
arises only in the second order in PHA. As it has been shown in
the main text, the above conclusions are also relevant to any
many-body corrections to thermomagnetic coefficients.

\section{AL Blocks}
For an arbitrary electron momentum relaxation time $\tau$, the AL
blocks ${\bf B}^{e,h,H}$ built from electron Green functions
$G^{R(A)}$ with electron vertices $\gamma$ ($ \gamma^e $, $
\gamma^H $, $ \gamma^h_1 $, and $ \gamma^h_2 $) are given by
\cite{20}
\begin{eqnarray}\label{B}
{\bf B}^{e,h,H}_i = {\rm Im}  \int {d {\bf p}\over (2 \pi)^n} {d
\epsilon \over 2 \pi} \ \gamma^{e,h,H}_i \ S_0(\epsilon) \
{(G_p^A)^2 G^R_{q-p} \over (1- \zeta)^2}, \ \
\\ \zeta =  {1\over
\pi \nu \tau} \int {d{\bf p} \over (2\pi )^3} \ G^A_p G^{R}_{q-p},
\ \ \ \ \ \ \ \ \ \ \ \ \ \ \ \ \ \  \ \ \ \ \ \ \
\end{eqnarray}
where the electron Green functions are given by Eq. A.2.

The block ${\bf B}^e$ with the electric current vertex, $\gamma^e=
e{\bf v}\cdot{\bf e}_E $, may be presented as \cite{20}
\begin{eqnarray}\label{Be}
{\bf B}^e({\bf q}) \ = \ 2e \ \nabla_{\bf q} P^R({\bf q},0)\cdot
{\bf e}_E= \ 2e\nu\alpha \ {\bf q}\cdot {\bf e}_E.
\end{eqnarray}
The block ${\bf B}^H$ with the vertex  $\gamma^H = (e/c){\bf
v}\cdot{\bf A}_H$ is given by \cite{20}
\begin{eqnarray}\label{BA}
{\bf B}^A({\bf q}) \ = \ {2e\over c} \ \nu\alpha \ {\bf q}\cdot
{\bf A}_H.
\end{eqnarray}

The block ${\bf B}^h_1$ with the kinetic heat current vertex,
$\gamma^h_1=
 \xi {\bf v}\cdot {\bf e}_{j^h} $ (${\bf e}_{j^h} =
 {\bf j}^h/j^h  \ \| {\bf A}$),
is given by \cite{27} (see also \cite{18,19,20})
\begin{eqnarray}\label{Bh1}
{\bf B}^h_1({\bf q},\omega) \ = \ \omega \ \nabla_{\bf q} P^R({\bf
q},0)\cdot {\bf e}_{j^h}= \omega \nu\alpha \ {\bf q}\cdot {\bf
e}_{j^h}.
\end{eqnarray}
Next,  we calculate the block ${\bf B}^h_2$ with the magnetic heat
current vertex $\gamma^h_2= ({\bf v} \cdot {\bf A}_H) ({\bf
v}\cdot {\bf e}_{j^h})$. The integral over angles of the electron
momentum involves only the vertex $\gamma^h_2$, because the heat
current is in the direction of ${\bf A}_H$. To obtain an imaginary
part in Eq. \ref{B}, the integral
\begin{eqnarray}
 \int d\xi \ (G_p^A)^2 G^R_{q-p}= {2 \pi i \over (2\epsilon-\omega
 -{\bf q}\cdot {\bf v}-i/\tau)^2},
\end{eqnarray}
should be expanded in $\omega$ (in calculations of ${\bf B}_e$ it
is expanded in ${\bf q}\cdot {\bf v}$). Finally, we get
\begin{eqnarray}\label{Bh2}
{\bf B}^h_2({\bf q}, \omega)  =  2 \omega  \nabla_{\bf q}^2
P^R({\bf q},0) A_H = 2 (e/c) \omega \nu \alpha A_H.
\end{eqnarray}

\section{Gauge ${\bf E}=i\Omega {\bf A}_E/c$}
Here we will show how our results can be obtained in the gauge,
where ${\bf E}=i\Omega {\bf A}_E/c$ and $ {\bf H}=i [ {\bf k}
\times {\bf A}_H ]$ (this was a question of one of our referees).
In the electric and magnetic fields the kinetic energy has a form
$ K = ({\bf p} + e{\bf A}/c)^2/2m $, and the part of the
Hamiltonian describing the interaction with external fields is
given by
\begin{eqnarray}\label{HH}
H'={e\over mc} {\bf p}( {\bf A}_H+{\bf A}_E) + {e^2\over 2mc^2} (
{\bf A}_H+{\bf A}_E)^2,
\end{eqnarray}
Calculating a response to ${\bf E} \times {\bf H}=(\Omega/c)[{\bf
A}_H({\bf k}\cdot{\bf A}_E)-{\bf k}({\bf A}_E\cdot{\bf A}_H]$, it
is convenient to use the gauge conditions ${\bf k}\cdot {\bf A}=0$
and ${\bf A}_H\cdot {\bf A}_E =0$  \cite{26}. In this gauge the
second term in Eq. \ref{HH} can be neglected. Including the
interaction with the magnetic field we get the heat current
operator,
\begin{eqnarray}\label{Jh}
{\hat {\bf J}}^h &=& \sum_{\bf p} {\bf v} \xi_p \ a_{\bf p}^+
a_{\bf p} + \sum_{\bf p}{ e{\bf v}\over c} ({\bf v}\cdot{\bf A}_H)
\ a_{\bf p}^+a_{\bf p}.
\end{eqnarray}
As it is expected, the term in the heat current describing the
interaction with the magnetic field is independent on the
presentation of the electric field (see Eqs. 5 and A1). Therefore,
all further calculations of thermomagnetic coefficient are the
same as in the gauge $E=-\nabla \phi$.

\begin{table*}
\caption{AL blocks (operators for fluctuating pairs) ${\bf B}$
based on electron operators ${\bf \gamma}$.}
\begin{ruledtabular}
\begin{tabular}{llll}
$ {\displaystyle \gamma^e = e \ {\bf v}\cdot{\bf e}_E }$ &
${\displaystyle \gamma^H = (e/c) \ {\bf v}\cdot{\bf A}}$ &
${\displaystyle \gamma^h_1= \xi \ {\bf v}\cdot {\bf e}_{j^h}}$ &
${\displaystyle \gamma^h_2= ({\bf v} \cdot {\bf A}_H) ({\bf
v}\cdot {\bf e}_{j^h})} $
\\[6pt]
 ${\displaystyle {\bf B}^e = 2e \ \nabla_{\bf q} P^R({\bf q},0)\cdot
{\bf e}_E }$ \ &
 ${\displaystyle {\bf B}^H = (2e/c) \ \nabla_{\bf q} P^R({\bf q},0)\cdot
{\bf A}_H} $ & ${\displaystyle {\bf B}^h_1= \omega \ \nabla_{\bf
q} P^R({\bf q},0)\cdot {\bf e}_{j^h}} $ & ${\displaystyle {\bf
B}^h_2=  \ 2 \omega  \ \nabla_{\bf q}^2 P^R({\bf q},0) } \ A_H
$
\\[2pt] $ {\displaystyle  \ \ \ \ \ =
 2e \nu \alpha \ {\bf q} \cdot {\bf e}_E} $ & ${\displaystyle  \ \ \ \ \ = (2e/c)\nu \alpha \
{\bf q}\cdot {\bf A}_H} $ & $ {\displaystyle  \ \ \ \ \ = \omega
\nu\alpha \ {\bf q}\cdot {\bf e}_{j^h}}$ & ${\displaystyle \ \ \ \
\ = (2e/c) \omega \nu \alpha \ {\bf A}_H \cdot {\bf e}_{j^h} }$
\\
\end{tabular}
\end{ruledtabular}
\end{table*}

\end{document}